\documentstyle[twocolumn,aps,psfig]{revtex}

\begin{document}
\draft

\twocolumn[\hsize\textwidth\columnwidth\hsize\csname @twocolumnfalse\endcsname

\title{Magnetization plateau in the spin ladder 
with the four-spin exchange}

\author{T\^oru Sakai and Yasumasa Hasegawa}
\address{
Faculty of Science, Himeji Institute of Technology, Kamigori,
Ako-gun, Hyogo 678-1297, Japan
}

\date{September 98}
\maketitle 

\begin{abstract}
The magnetization process of the $S$=1/2 antiferromagnetic spin ladder  
with the four-spin cyclic exchange interaction at $T=0$ 
is studied by the exact
diagonalization of finite clusters and size scaling analyses. 
It is found that a magnetization plateau appears at half the saturation
value if the ratio of the four- and two-spin exchange coupling constants
$J_4$ 
is larger than the critical value $J_{4c}=$0.05$\pm$0.04.  
The phase transition with respect to $J_4$ 
at $J_{4c}$ is revealed to be 
the Kosterlitz-Thouless-type. 
\end{abstract}

\vskip2pc]
\narrowtext

%
%

The field-induced spin gap is one of recent interesting topics 
on the one-dimensional (1D) quantum spin systems. 
The gap can be detected as a plateau of the magnetization curve 
at low temperatures. 
The appearance of such a magnetization plateau 
was theoretically predicted in several systems; 
the anisotropic $S={3\over 2}$ 
antiferromagnetic chain\cite{oshikawa,s3-2}, 
the $S=1$ bond-alternating chain\cite{tonegawa}, 
the $S={1\over 2}$ ferromagnetic-ferromagnetic-antiferromagnetic chain
\cite{ffa}, 
the frustrated bond-alternating chain\cite{totsuka}, 
the three-leg ladder\cite{cabra}, 
and the frustrated two-leg ladder\cite{mila}. 
In particular the two-leg ladder attracts a great interest 
in the context of the superconductivity in a doped system\cite{rice}. 
The standard $S={1\over 2}$ uniform antiferromagnetic spin ladder 
has the spin gap of the lowest excitation from the nonmagnetic ground 
state (GS), which leads to a plateau at the magnetization 
$m=0$\cite{hida,dagotto,troyer}. 
A strong coupling approach\cite{mila} indicated an additional plateau 
at half the saturated magnetization due to 
the next-nearest antiferromagnetic coupling which yields 
the frustration. 
In this paper we show another possibility of the magnetization 
plateau in the two-leg spin ladder, 
which is induced by a four-spin exchange interaction.   

%
%
 
A four-spin exchange interaction described by a product of two-spin 
exchanges in a spin ladder 
was investigated by a field theoretical approach\cite{tsvelik}.  
It indicated the possibility of a different type of massive phase 
from the Haldane phase\cite{haldane}
in the nonmagnetic GS, 
but the state in a strong magnetic field was not discussed.  
On the other hand 
a mean field analysis\cite{kubo} suggested that 
the $S={1\over 2}$ triangular lattice antiferromagnet 
would have a magnetization plateau at half the saturated magnetization,
if there exists a four-spin cyclic exchange interaction. 
It was verified by the exact diagonalization\cite{misguich}. 
The recent experiments 
revealed that such multiple-spin exchange 
interactions are realized in the two-dimensional (2D)
solid $^3$He\cite{he1,he2} and the 2D 
Wigner solid of electrons formed 
in a Si inversion layer\cite{wigner}, 
as well as the bcc $^3$He\cite{he3}.  
In order to test the possibility of a similar magnetization plateau 
in 1D quantum spin systems, 
we consider the $S={1\over 2}$ uniform antiferromagnetic spin ladder 
with the four-spin cyclic exchange at every plaquette. 
The magnetization process of the system is described by the Hamiltonian 
\begin{eqnarray}
\label{ham}
&{\cal H}&={\cal H}_0+{\cal H}_Z, \nonumber \\
&{\cal H}_0&=\sum_j({\bf S}_{1,j}\cdot {\bf S}_{1,j+1}
               +{\bf S}_{2,j}\cdot {\bf S}_{2,j+1}
               +{\bf S}_{1,j}\cdot {\bf S}_{2,j})
 \nonumber \\
&&   +  J_4 \sum _j (P_{4,j}+P^{-1}_{4,j}) 
, \\
&{\cal H}_Z& =-H\sum _j (S_{1,j}^z+S_{2,j}^z), \nonumber
\end{eqnarray}
where $P_{4,j}$ is the cyclic permutation operator which 
exchanges the four spins around the $j$-th plaquette as 
${\bf S}_{1,j}\rightarrow {\bf S}_{1,j+1}\rightarrow
{\bf S}_{2,j+1}\rightarrow {\bf S}_{2,j}\rightarrow {\bf S}_{1,j}$, 
$J_4$ is the strength of the four-spin exchange and $H$ is the 
applied magnetic field normalized by the two-spin exchange coupling 
constant. 
We assume $J_4$ is positive, as it is in the 
solid $^3$He.  
This system subjected to the periodic boundary condition is 
studied by the exact diagonalization of the finite clusters and 
the size scaling of the low-lying energy spectra. 
For $L\times 2$-spin systems, 
the lowest energy of ${\cal H}_0$ in the subspace where 
$\sum _j (S_{1,j}^z+S_{2,j}^z)=M$ 
is denoted as $E(L,M)$. 
Using Lanczos' algorithm, we calculated $E(L,M)$ 
($M=0,1,2,\cdots,L$) for even-$L$ systems up to $L=16$. 
The macroscopic magnetization is defined as $m={M\over L}$.  

%
%

The nonmagnetic GS 
of the system (\ref{ham}) with $J_4=0$ is in a massive
phase equivalent to the Haldane phase of the $S=1$ antiferromagnetic
chain and the low-lying excitation has a finite energy gap for $m=0$. 
On the other hand, 
the magnetic GS is always gapless
\cite{hayward,chitra} except for the saturation. 
Thus the magnetization curve has a plateau at $m=0$, 
while no other plateau appear up to $m=1$, as far as $J_4=0$. 
The four-spin exchange, however, is expected to induce a plateau 
at $m={1\over 2}$, because the interaction stabilizes 
the '{\it uuud}' state, mentioned in Ref. \cite{kubo}, 
of the four spins around every plaquette within a mean field argument.
We concentrate on the plateau at $m={1\over 2}$, 
rather than the nonmagnetic GS. 

The magnetic excitation gap giving $\delta M =\pm 1$ of 
the $L\times 2$-spin systems 
described by the total Hamiltonian ${\cal H}$ is given by 
\begin{eqnarray}
\label{gap}
\Delta _{\pm} \equiv E(L,M\pm 1)-E(L,M) \mp H. 
\end{eqnarray}
For the gapless system in the thermodynamic limit, 
the conformal field theory\cite{cft} (CFT) 
predicts the asymptotic form of the size
dependence of the gap as $\Delta _{\pm} \sim O(1/L)$ 
with fixed $m=M/L$. 
When $H_+$ and $H_-$ are defined as 
\begin{eqnarray}
\label{field1}
E(L,M+1)-E(L,M) \rightarrow H_+ \quad (L \rightarrow \infty), \nonumber \\
E(L,M)-E(L,M-1) \rightarrow H_- \quad (L \rightarrow \infty),
\end{eqnarray}
$H_+$ and $H_-$ has the same value
and it gives the magnetic field $H$ for the magnetization $m$ in the 
thermodynamic limit. 
In contrast to the gapless case, 
if the system has a finite gap even in the infinite length limit, 
$\Delta _+$ and $\Delta _-$ are still finite for $L \rightarrow \infty$.
It leads to the difference between   
$H_+$ and $H_-$ 
and  
a plateau appears for $H_- < H < H_+$ at $m=M/L$ in the 
magnetization curve at $T=0$. 

%
%
\begin{figure}[htb]
\begin{center}
\mbox{\psfig{figure=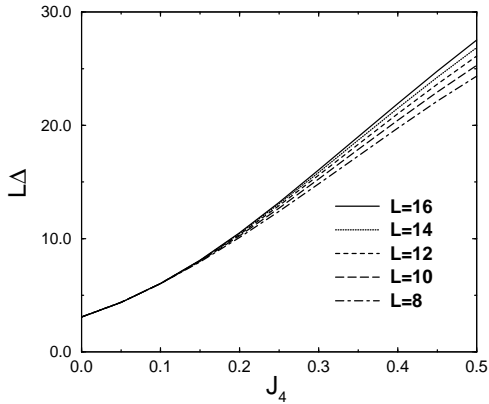,width=7cm,height=6cm,angle=-90}}
\end{center}
\caption{
Scaled gap $L\Delta$ versus the strength of the four-spin exchange 
interaction $J_4$.
\label{fig1}
}
\end{figure}
%
%
\begin{figure}[htb]
\begin{center}
\mbox{\psfig{figure=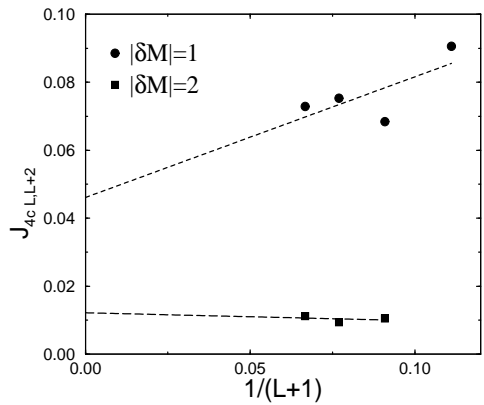,width=7cm,height=6cm,angle=-90}}
\end{center}
\caption{
$L$-dependent fixed point $J_{4c L,L+2}$ of the gap for $\delta M=\pm 1$
(circles) and $\delta M=\pm 2$ (squares) 
are plotted 
versus $1/(L+1)$ to determine $J_{4c}$ in the thermodynamic limit. 
The estimated value is $J_{4c}=0.05 \pm 0.01$ for $\delta M=\pm 1$, 
which does not well agree with the result for $\delta M=\pm
2$ $J_{4c}=0.01 \pm 0.01$. 
Thus we conclude $J_{4c}=0.05 \pm 0.04$.  
\label{fig2}
}
\end{figure}

The sum $\Delta \equiv \Delta _+ +\Delta _-$ is a good order 
parameter to investigate the plateau-nonplateau transition 
with the finite-size scaling\cite{s3-2}, 
because $\Delta$ corresponds to the length of the plateau in the
magnetization curve in the thermodynamic limit. 
The scaled gap $L\Delta$ of finite systems ($L=8 \sim 16$) at $m=1/2$ 
is plotted versus $J_4$ in Fig. \ref{fig1}. 
For $J_4>0.2$ the scaled gap obviously increases with increasing $L$, 
which means that a finite gap exists in the thermodynamic limit. 
For small $J_4$ around the region $0<J_4<0.1$, 
the scaled gap looks almost independent of $L$. 
It implies that the system is gapless at a finite region of 
the parameter $J_4$, 
which is reminiscent of the Kosterlitz-Thouless (KT)
transition\cite{kt}.  
According to our precise analysis,  
the $L\Delta$ curves for $L$, and $L+2$ have an intersection 
in the region $0<J_4<0.1$ for each $L$. 
Thus the critical point $J_{4c}$ can be estimated by the phenomenological
renormalization group (PRG) equation\cite{phenomenological}
\begin{eqnarray}
\label{prg}
(L+2)\Delta_{L+2}(J_4')=L\Delta_L(J_4).
\end{eqnarray}
We define $J_{4c L,L+2}$ as the $L$-dependent fixed point of (\ref{prg}) and 
it is extrapolated to the thermodynamic limit. 
$J_{4c L,L+2}$ is plotted versus $1/(L+1)$ as solid circles in Fig. \ref{fig2}. 
Although the convergence of $J_{4c L,L+2}$ with increasing $L$ is not
good, 
the least square fitting of the form $J_{4c L,L+2} \sim J_{4c}+A/(L+1)$ 
gives the extrapolated result $J_{4c}=0.05 \pm 0.01$ as the dashed line 
in Fig. \ref{fig2}. 
To test the precision of the value, 
we did the same analysis using the gap for $\delta M=\pm 2$ instead of
$\Delta _{\pm}$ as solid squares shown in Fig. \ref{fig2} 
where the fixed point can be obtained only for $L \geq 10$.   
It gave $J_{4c}=0.01 \pm 0.01$ which is not well coincide with the above 
result, which implies that the available system size is not enough 
to determine $J_{4c}$ with the fitting of $1/(L+1)$. 
(Such a difficulty of the precise decision of the critical point by 
PRG is sometimes due to the 
logarithmic size correction in the case of the KT 
transition.\cite{nomura})
Assuming that the system is gapless for $J_4=0$, 
we conclude $J_{4c}=0.05 \pm 0.04$ within the present analysis. 

%
%
We present the GS magnetization curve in the 
thermodynamic limit for $J_4$=0 and 0.1. 
In the latter case the magnetization plateau should appear at $m={1\over
2}$ in contrast to the former, as discussed above. 
Note that the four-spin exchange interaction reduces the spin gap 
just above the nonmagnetic GS. 
According to our present analysis based on PRG,  
the gap for $m=0$ 
vanishes at a critical value ${\tilde J}_{4c}$, 
which should be distinguished from $J_{4c}$ for $m={1\over 2}$, 
and the nonmagnetic GS will belong to a different massive phase 
from the Haldane phase for $J_4 > {\tilde J}_{4c}$. 
The critical value ${\tilde J}_{4c}$, however, 
is obviously larger than 0.1. 
Thus even in the case of $J_4=0.1$ 
the spin gap due to the Haldane mechanism still exists 
for $m=0$ and we can use the same method to give the magnetization 
curve as used for the $S=1$ antiferromagnetic chain\cite{sakai}.  

\begin{figure}[htb]
\begin{center}
\mbox{\psfig{figure=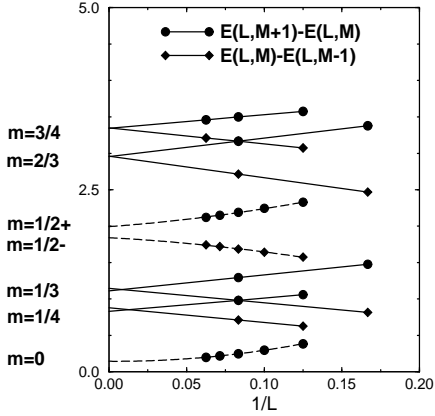,width=7cm,height=6cm,angle=-90}}
\end{center}
\caption{
$E(L,M+1)-E(L,M)$ and $E(L,M)-E(L,M-1)$ plotted 
versus $1/L$ with fixed $m$ for $J_4=0.1$. 
The dashed curves are guides to the eye. 
The extrapolated points for $m=0$, $m=1/2-$ and $m=1/2+$ correspond to 
the results of the Shanks' transformation $H_{c1}=0.15 \pm 0.03$, 
$H_-=1.84 \pm 0.06$ and $H_+=1.99 \pm 0.09$, 
respectively.  
\label{fig3}
}
\end{figure}

For $J_4=0.1$ 
the left hand sides of the form (\ref{field1}) calculated for 
$m=0,{1\over 4}, {1\over 3}, {1 \over 2}, {2\over 3}$ and ${3\over 4}$
are plotted versus $1/L$ in Fig. \ref{fig3}.  
It shows $H_+=H_-$ except for $m=0$ and ${1\over 2}$. 
Thus we take the mean value of the two for the magnetic field $H$ 
for each $m$.  

%
%

Since the nonmagnetic GS is massive for $J_4=$0 and 0.1, 
the size correction of $H_+$ decays faster than ${1\over L}$   
as shown in Fig. \ref{fig3}.
Thus we use the Shanks' transformation\cite{shanks}
$P'_n=(P_{n-1}P_{n+1}-P_n^2)/(P_{n-1}+P_{n+1}-2P_n)$
twice for the sequence $E(L,1)-E(L,0)$ 
for $L=$6,8,10,12 and 14,
and obtain $H_{c1}=0.503 \pm 0.003$ and $0.15 \pm 0.03$ for 
$J_4$=0 and 0.1, respectively. 
The saturation field $H_{c2}$ is given by 
the $L$-independent quantity $E(L,L)-E(L,L-1)$. 

%
%
\begin{figure}[htb]
\begin{center}
\mbox{\psfig{figure=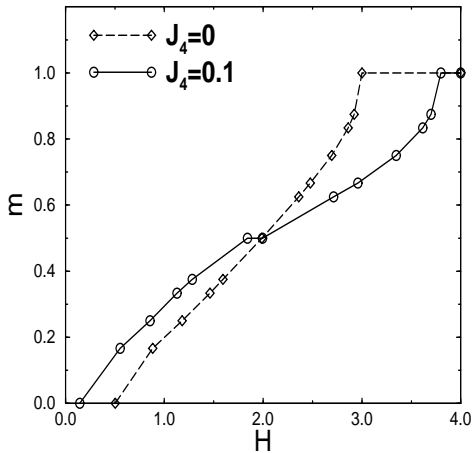,width=7cm,height=7cm,angle=-90}}
\end{center}
\caption{
GS magnetization curves in the thermodynamic limit 
for $J_4=0$ and 0.1. 
The latter has the magnetization plateau at half the saturation value. 
\label{fig4}
}
\end{figure}

In the case of $J_4=0.1$,  
for $m=1/2$ $H_+$ and $H_-$ are obviously different 
and the size correction decays faster than $1/L$,  
as shown in Fig. \ref{fig3}, 
which is consistent with a finite gap. 
Then we estimate $H_+$ and $H_-$ 
by the Shanks' transformation  
and get $H_+=1.99 \pm 0.09$ and $H_-=1.84\pm 0.06$.  
For $J_4=0$ 
$H_+$ and $H_-$ correspond even at $m=1/2$.  
We present the results for $J_4=$0 and 0.1 in Fig. \ref{fig4}, 
where we also used the values of $H$ for $m=$1/6, 3/8, 5/8, 5/6 and 7/8 
which are estimated by the same method as mentioned above. 
The curve has a plateau at $m=1/2$ ($H_- <H<H_+$) for $J_4=$0.1.   

%
%

Our present PRG analysis shows that 
the the gap does not behaves as $\Delta \sim(J_4-J_{4c})^{\nu}$. 
If we define the the size-dependent exponent $\nu _L$, 
it diverges as $L$ increases. 
Instead, if the gap behaves like 
\begin{eqnarray}
\label{ktgap}
\Delta \sim \exp \big( -{ a \over {(J_4-J_{4c})^{\sigma}}} \big),
\end{eqnarray}
as in the case of universality class 
of KT transitions $(\sigma ={1\over 2})$, 
the Roomany-Wyld 
approximation for the Callen-Symanzik $\beta$-function\cite{rw}, 
which is defined as 
\begin{eqnarray}
\label{beta1}
\beta _{L,L+2} (J_4) = {{ 
1+\log\big({{\Delta_{L+2}(J_4)}\over {\Delta_L(J_4)}} \big) \big/
\log \big({{L+2}\over L}\big) }\over{
\big[{{\Delta '_L(J_4)\Delta '_{L+2}(J_4)}\over {\Delta_L(J_4) \Delta_{L+2}
(J_4)}}\big]^{1\over 2}}},
\end{eqnarray}
should have the form 
\begin{eqnarray}
\label{beta2}
\beta _{L,L+2} (J_4) \sim (J_4-J_{4c L,L+2})^{1+\sigma}. 
\end{eqnarray}
Fitting the form (\ref{beta2}) to the calculated function
(\ref{beta1}) for each $L$, 
$\sigma$ is estimated as follows: 
$\sigma _{10,12}=0.38 \pm 0.10$, $\sigma _{12,14}=0.43 \pm 0.10$ and 
$\sigma _{14,16}=0.49 \pm 0.10$. 
The results are consistent with $\sigma ={1\over 2}$. 
Thus we conclude the critical behavior near $J_{4c}$ for $m={1\over 2}$ 
is characterized by the universality class of the 
KT transition. 

\begin{figure}[htb]
\begin{center}
\mbox{\psfig{figure=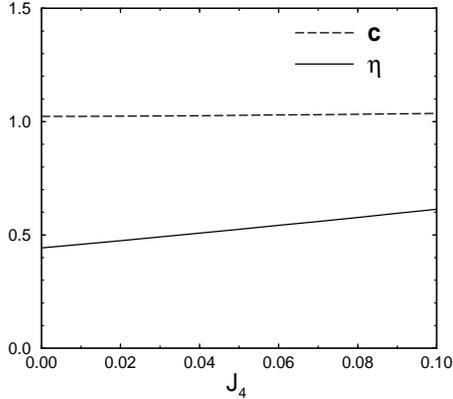,width=7cm,height=6cm,angle=-90}}
\end{center}
\caption{
Estimated central charge $c$ and exponent $\eta$ around $J_{4c}$.  
The result indicates $c=1$ which is consistent with  
the KT transition. 
$\eta$ is close to ${1\over 2}$ rather than ${1\over 4}$. 
\label{fig5}
}
\end{figure}

Furthermore we estimate the central charge $c$ of CFT,  
using the 
asymptotic form of 
the GS energy per site 
\begin{eqnarray}
\label{gs}
{1\over L}E(L,M) \sim \epsilon (m) -{{\pi} \over 6}cv_s {1\over {L^2}}
\qquad (L \rightarrow \infty),
\end{eqnarray}
where $v_s$ is the sound velocity which is the gradient of the
dispersion curve at the origin. 
The result shown in Fig. \ref{fig5} 
suggests $c=1$ with only a few percent errors for 
$m={1\over 2}$ and $0\leq J_4 \leq 0.1$. 
It also supports the KT transition.

The critical exponent $\eta$, associated with the spin correlation
function in the leg direction 
like $\langle S_0^+ S_r^- \rangle \sim (-1)^r r^{-\eta}$, 
can be estimated by the form of the gap 
$\Delta _{\pm} \sim \pi v_s \eta/L \quad (L\rightarrow \infty)$
\cite{s3-2}. 
The estimated $\eta$, shown in Fig. \ref{fig5}, 
seems close to ${1\over 2}$ around the critical point $J_{4c}$, 
rather than ${1\over 4}$ which is expected for the KT 
transition. 
We think there is a possible jump 
from $\eta ={1\over 4}$ to $\eta ={1\over 2}$ at $J_{4c}$  
because the elementary excitations is expected to behave 
like the free Fermion systems ($\eta ={1\over 2}$) 
at the edge of the plateau for $J_4>J_{4c}$\cite{totsuka}.  
The present small cluster analysis could not detect such a
discontinuity. 
Another exponent $\eta ^z$ defined as 
$\langle S_0^z S_r^z \rangle \sim \cos (2k_Fr) r^{-\eta^z}$ 
can also be estimated from the $L$-dependence of the soft mode gap 
with the momentum $2k_F=2\pi m$\cite{chitra}.  
We checked the validity of the relation $\eta \eta ^z=1$ around 
$J_{4c}$ which is consistent 
with the Luttinger liquid theory\cite{luttinger} 
leading to $\eta ={1\over 2}$ in the free Fermion case. 

%
%

The spin gap at $m=0$ has already been observed in several real 
ladder compounds, for example Cu$_2$(C$_2$H$_{12}$N$2$)$_2$Cl$_4$
\cite{experiment,hayward} 
and La$_6$Ca$_8$Cu$_{24}$O$_{41}$\cite{imai}.  
The magnetization plateau, however, has not been detected 
at any finite magnetization. 
We hope some new ladder materials 
with the field-induced spin gap will be discovered in the near future.

%
%
In summary 
the finite cluster calculation and size scaling study showed that 
the $S={1\over 2}$ antiferromagnetic spin ladder with the 
four-spin cyclic exchange interaction at every plaquette 
has the magnetization plateau 
at $m=1/2$ for $J_4>J_{4c}=0.05 \pm 0.04$ and 
the phase transition with respect to $J_4$ belongs to 
the KT universality class.   

%
%

We wish to thank Prof. K. Nomura for fruitful 
discussions. 
We also thank 
the Supercomputer Center, Institute for
Solid State Physics, University of Tokyo for the facilities
and the use of the Fujitsu VPP500.

\end{document}